\documentclass[conference,a4paper]{IEEEtran}
\usepackage[left=1.34cm,right=1.34cm,top=1.91cm,bottom=4.38cm]{geometry}
\usepackage{balance}
\usepackage{float}

\usepackage{graphicx}
\usepackage{subfig}
\usepackage{tabularx}

\usepackage{standard}
\usepackage{etex}

\usepackage{rotate}
\usepackage{algorithmic}
\usepackage{algorithm}
\usepackage{float}
\usepackage{tikz}
\usepackage{refcount}

\usepackage{pgf,tikz}
\usepackage{pgfplots}
\usepackage{pgfplotstable}
\usepackage{bbm}
\usepackage{adjustbox}
\usetikzlibrary{calc}
\usepackage{relsize}
\usepackage{ifthen}

\usetikzlibrary{calc,fit,arrows,plotmarks,intersections,patterns,shapes,decorations,positioning, backgrounds, matrix}

\usepackage[numbers,sort&compress]{natbib}
\usepackage[ngerman, english]{babel}

\usetikzlibrary{arrows.meta}
\edef\wdArrowLength{2}
\tikzset{>={Latex[width=1.5mm,length=\wdArrowLength mm]}}
\usepackage{graphicx}
\usepackage[outdir=./pictures]{epstopdf}

\usepackage{cancel}
\usepackage{mathtools}
\usepackage{shadethm}
\usepackage{empheq}
\usepackage{skull}
\usepgfplotslibrary{units}
\usepackage{calc}
\usepackage{nicefrac}
\usepackage{fancybox}
\usepackage{psfrag}
\usepackage{dsfont}

\title{Design and Deployment Guidelines for UAV-Mounted RIS Under Position Uncertainty
\thanks{This work was funded by the German Research Foundation (DFG) in the course of the project SPP2433 under the project no. 541021107 (Measurement Technology on Flying Platforms) under grant SE 1697/22-1,  grant MO 1086/17-1, and grant INST 213/1028-1 FUGB.}} 
\author{\author{Kevin Weinberger$^{\dagger }$, David Müller$^{\ast }$, Martin M\"onnigmann$^{\ast }$, Aydin Sezgin$^{\dagger }$, \\
        $^{\dagger }$Institute of Digital Communication Systems, Ruhr-Universit\"at Bochum, Germany,\\
        $^{\ast }$Automatic Control and Systems Theory, Ruhr-Universit\"at Bochum, Germany,\\
		 \{Kevin.Weinberger, David.Mueller-r21, Martin.Moennigmann, Aydin.Sezgin\}@rub.de  }}
\date{\today}

\usepackage{graphicx}
\usepackage{placeins}

\usepackage{booktabs}
\usepackage{marvosym}
\usepackage{multirow}

\usepackage{latexsym, amsmath, amssymb, amsfonts, upgreek}

\usepackage[nolist]{acronym}

\usepackage{tikz}
\usetikzlibrary{calc,arrows,positioning,decorations,shadows,shapes,fadings,matrix}
\tikzset{>=latex'}
\tikzset{semithick}

\makeatletter
\providecommand{\IfElsePackageLoaded}[3]{\@ifpackageloaded{#1}{#2}{#3}}
\makeatother

\makeatletter

\def\tikz@delimiter#1#2#3#4#5#6#7#8{%
	\bgroup
		\pgfextra{\let\tikz@save@last@fig@name=\tikz@last@fig@name}%
		node[outer sep=0pt,inner sep=0pt,draw=none,fill=none,anchor=#1,at=(\tikz@last@fig@name.#2),#3]
		{%
			{\nullfont\pgf@process{\pgfpointdiff{\pgfpointanchor{\tikz@last@fig@name}{#4}}{\pgfpointanchor{\tikz@last@fig@name}{#5}}}}%
			\delimitershortfall\z@
			\resizebox*{!}{#8}{$\left#6\vcenter{\hrule height .5#8 depth .5#8 width0pt}\right#7$}%
		}
		\pgfextra{\global\let\tikz@last@fig@name=\tikz@save@last@fig@name}%
	\egroup%
}

\makeatletter
\renewenvironment{thebibliography}[1]{%
\@xp\section\@xp*\@xp{\refname}%
\normalfont\footnotesize\labelsep .5em\relax
\renewcommand\theenumiv{\arabic{enumiv}}\let\p@enumiv\@empty
\vspace*{-2pt}
\list{\@biblabel{\theenumiv}}{\settowidth\labelwidth{\@biblabel{#1}}%
\leftmargin\labelwidth \advance\leftmargin\labelsep
\usecounter{enumiv}}%
\sloppy \clubpenalty\@M \widowpenalty\clubpenalty
\sfcode`.=\@m
}{%
\def\@noitemerr{\@latex@warning{Empty `thebibliography' environment}}%
\endlist
}
\makeatother

\let\OLDthebibliography\thebibliography
\renewcommand\thebibliography[1]{
\OLDthebibliography{#1}
\setlength{\parskip}{-0.5pt}
\setlength{\itemsep}{-0.1pt}
}

\tikzset{hexagon/.code={
	\draw (0,2) -- (-4,0) -- (0,-2) -- (4,0) -- (0,2);
}}

\tikzset{phone/.code={
   \node [rectangle,rounded corners=1.5pt,draw,minimum height=0.6cm, minimum width=0.35cm] at (0,0){};
   \node [rectangle,rounded corners=1.5pt,draw,minimum height=0.5cm, minimum width=0.3cm] at (0,0){};
}}

\makeatletter
\def\cantox@vector#1#2#3#4#5#6#7#8{%
  \dimen@.5\p@
  \setbox\z@\vbox{\boxmaxdepth.5\p@
   \hbox{\kern-1.2\p@\kern#1\dimen@$#7{#8}\m@th$}}%
  \ifx\canto@fil\hidewidth  \wd\z@\z@ \else \kern-#6\unitlength \fi
  \ooalign{%
    \canto@fil$\m@th \CancelColor
    \vcenter{\hbox{\dimen@#6\unitlength \kern\dimen@
      \multiply\dimen@#4\divide\dimen@#3 \vrule\@depth\dimen@\@width\z@
      \vector(#3,-#4){#5}%
    }}_{\raise-#2\dimen@\copy\z@\kern-\scriptspace}$%
    \canto@fil \cr
    \hfil \box\@tempboxa \kern\wd\z@ \hfil \cr}}
\def\bcancelto#1#2{\let\canto@vector\cantox@vector\cancelto{#1}{#2}}
\makeatother

\newcommand{\ifthen}[2]{\ifthenelse{#1}{#2}{}}

\definecolor{myblue1}{rgb}{0,0,255}
\definecolor{myblue2}{rgb}{65,105,225}
\definecolor{myblue3}{rgb}{70,130,180}
\definecolor{myblue4}{rgb}{176,196,222}

\newcommand{\mytilde}{{\raise.17ex\hbox{$\scriptstyle\mathtt{\sim}$}}}
\newcommand{\naive}{}
\def\naive/{na\"{\i}ve}

\newcommand{\executeiffilenewer}[3]{%
\ifnum\pdfstrcmp{\pdffilemoddate{#1}}%
{\pdffilemoddate{#2}}>0%
{\immediate\write18{#3}}\fi%
}

\pgfkeys{/pgf/fpu}
\pgfmathparse{16383+1}

\pgfkeys{/pgf/fpu=false}

\IEEEoverridecommandlockouts
\usepackage{pdfpages}

\begin{document}
\bstctlcite{IEEEexample:BSTcontrol}

\maketitle

\begin{abstract}
UAV-mounted reconfigurable intelligent surfaces (RIS) are a promising enabler for 6G networks, offering dynamic control of wireless propagation for coverage enhancement, integrated sensing and communication (ISAC), and localization. By exploiting UAV mobility, RIS can maintain favorable line-of-sight links, improving channel quality in dynamic environments. However, UAV positioning uncertainties introduce channel distortions that degrade RIS phase alignment and coherent combining. This work develops a GUM-based uncertainty propagation framework for UAV-mounted RIS channels, mapping UAV position uncertainty through the geometric Tx–RIS–Rx model into the complex cascaded channel. We derive a closed-form stochastic propagation model capturing nonlinear phase uncertainty effects and quantify their impact on channel coherence. The results show that phase uncertainty induces exponential coherence loss, dominating performance degradation. To characterize this transition, we introduce a performance-driven coherence threshold (PCT) that defines the boundary where incoherent combining results in a predetermined performance loss. Results based on analytical scaling laws and Monte Carlo simulations confirm the tightness of the PCT in accurately capturing the coherence transition. This validated threshold is then leveraged to derive optimal UAV-mounted RIS placement, revealing that realistic positioning conditions significantly deviate from the conventional RIS intuition, which typically favors placement close to either the transmitter or receiver.
\end{abstract}
	\vspace{-0.1cm}
\section{Introduction}
Future 6G wireless networks are expected to support highly dynamic environments with integrated connectivity, sensing, and localization \cite{6Gvision}. In this context, reconfigurable intelligent surfaces (RIS) have emerged as a key enabler due to their ability to programmatically control electromagnetic wave propagation in a cost- and energy-efficient manner, improving coverage, spectral efficiency \cite{6G_RIS}, resilience \cite{weinberger2025symbolcountsresilientwireless}, and enabling integrated sensing and communication (ISAC) \cite{AirToAirChannelshowmetheway}. To enhance flexibility, the concept of UAV-mounted RIS has recently attracted significant attention \cite{UAV-RIS}, as it enables controllable three-dimensional mobility.
 This is particularly attractive in scenarios with limited infrastructure, such as disaster recovery or rapidly changing networks, where UAV-mounted RIS can maintain favorable line-of-sight conditions and improve link quality compared to fixed deployments \cite{UAV_Kev}.

However, mounting an RIS on a UAV exposes the system to both external and internal sources of uncertainty, such as wind disturbances and sensor noise, which result in orientation \cite{UncertUAVMeas} and position errors \cite{Pos_uncert}. These directly affect Tx–RIS–Rx alignment and degrade phase-coherent combining, since RIS operation relies on precise phase control over large apertures. In contrast, most existing works assume perfect UAV positioning or adopt simplified additive uncertainty models, failing to capture the nonlinear impact of localization errors on RIS coherence. This leaves a gap in physically grounded models linking localization errors to coherent gain degradation.

To address this, we develop a GUM-based uncertainty propagation framework for UAV-mounted RIS channels, mapping UAV position uncertainty through the geometric Tx–RIS–Rx model into the complex cascaded channel. We derive a closed-form stochastic model capturing nonlinear phase uncertainty effects. As it turns out, amplitude uncertainty leads to mild additive power variations, whereas phase uncertainty induces exponential coherence loss that dominates large-scale RIS performance.

Motivated by this insight, we introduce a performance-driven coherence threshold (PCT) that characterizes the transition between coherent and incoherent phase combining as a function of RIS size, frequency, and uncertainty level, providing a compact design metric for RIS operation under uncertainty.
Finally, analytical scaling laws and measurement-informed Monte Carlo simulations validate the proposed PCT and reveal that optimal UAV-mounted RIS placement is jointly governed by geometry and uncertainty, yielding unintuitive design implications.

\section{System Model}\label{sec:sysmod}
%
%
\begin{figure}[]
	\centering
	\includegraphics[trim={2.2cm 1cm 0.25cm 1.5cm},clip,width=0.66\linewidth] {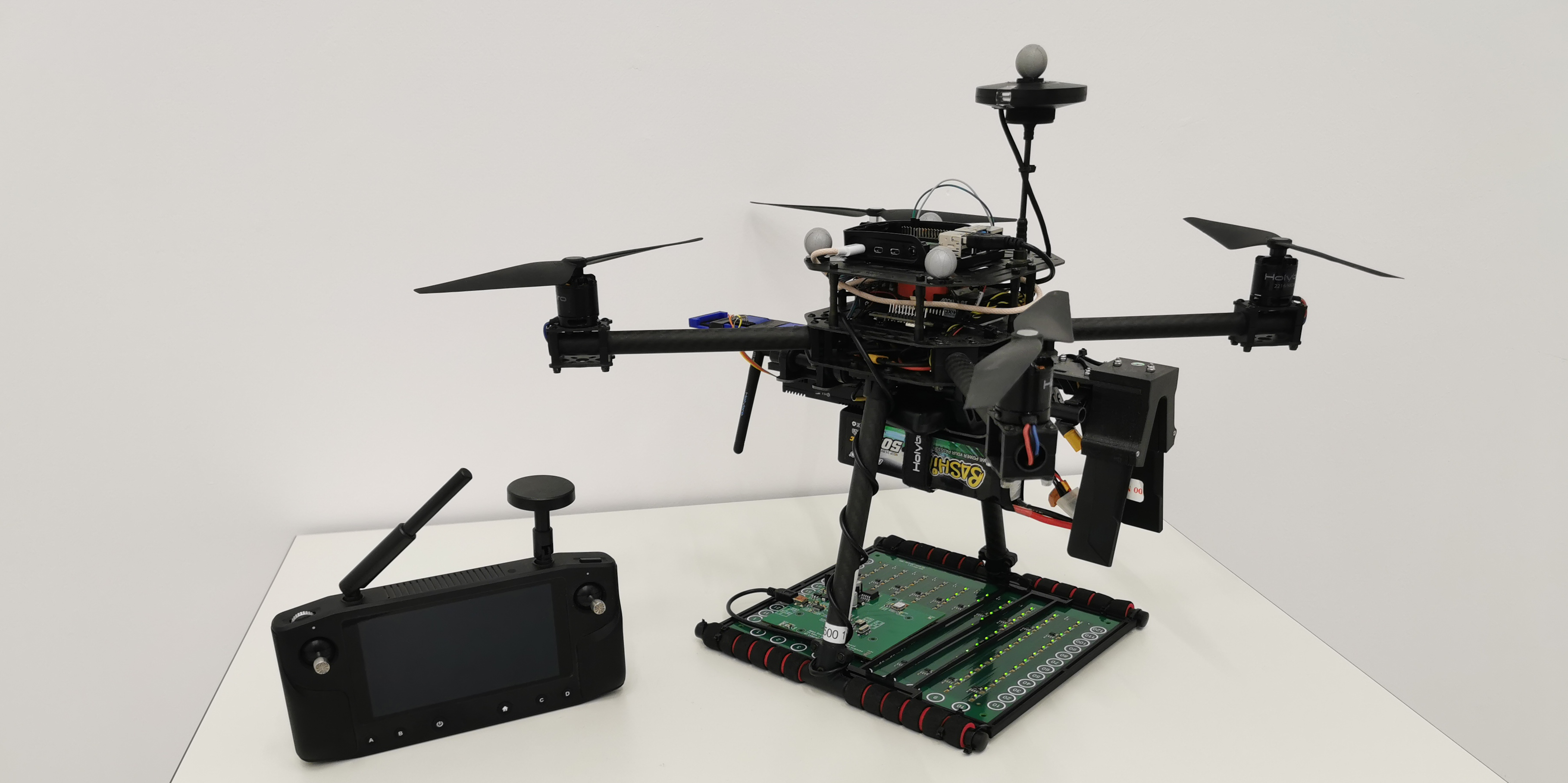} 
	\caption{\small Customized Holybro X500 featuring a mounted RIS prototype, controlled by a Raspberry Pi 4B, and Herelink 1.1 controller.}
	\label{fig:UAV}\vspace{-0.55cm}
\end{figure}
The UAV carrying the RIS (depicted in Fig.~\ref{fig:UAV}) is assumed to hover in the air while carrying the reflecting surface, which is oriented toward the ground. A single-antenna transmitter (Tx) sends signals to the RIS prototype \cite{OpenRIS}, which consists of \(M=120\) reflecting elements. The reflected signals are then received by a single-antenna receiver (Rx). By applying controllable phase shifts to the reflected waves, the resulting effective Tx-RIS-Rx channel is expressed as
\vspace{-0.25cm}
\begin{align}\label{eq:heff}
    h^\mathsf{eff}
    =
    \sqrt{G_T}\sqrt{G_R}
    \sum_{m=1}^{M}
    h_m \phi_m g_m,
    \\[-20pt]\nonumber
\end{align}
where \(h_m\) and \(g_m\) denote the channels between the Tx and the \(m\)-th RIS element, and between the \(m\)-th RIS element and the Rx, respectively. The antenna gains are represented by \(G_T\) and \(G_R\), while the RIS-induced phase shift is given by
\(
\phi_m = e^{j\varphi_m}
\),
with \(\varphi_m\) denoting the applied phase shift. To optimize the RIS phase configuration, knowledge of the cascaded Tx-RIS-Rx channels is required. Here, we exploit the instantaneous geometry of the setup. As shown in our previous work \citep{AirToAirChannelshowmetheway}, this geometry-based approach enables real-time RIS optimization with performance close to experimentally obtained configurations. Let \(d_m^h\) and \(d_m^g\) denote the distances from the Tx and Rx to the \(m\)-th RIS element, respectively. The cascaded channel component is then modeled as \cite{AirToAirChannelshowmetheway}
\vspace{-0.1cm}
\begin{align}\label{h_casc}
    h^\mathsf{casc}_m
    =
    h_m g_m
    =
    \left[
    \frac{c}{4\pi \nu d_m^h}
    e^{j\frac{2\pi}{\lambda}d_m^h}
    \right]
    \left[
    \frac{c}{4\pi \nu d_m^g}
    e^{j\frac{2\pi}{\lambda}d_m^g}
    \right],
\end{align}
where \(c\) is the speed of light, \(\nu\) is the carrier frequency, and \(\lambda\) denotes the wavelength.

\section{Uncertainty of the UAV's position}\label{sec:exp}
\subsection{Measurement Data Acquisition}

In this work, we use the UAV-mounted RIS setup and data introduced in~\cite{Pos_uncert}. The setup consists of a UAV-mounted RIS prototype, where position and orientation are estimated onboard using an EKF implemented on a Pixhawk Cube Orange flight controller. Ground truth data for uncertainty analysis was recorded using a high-precision motion capture system (MCS)\footnote{https://www.vicon.com/}. The MCS provides submillimeter accuracy and serves as the basis for uncertainty characterization.

A total of 10 hover flights were conducted under identical outdoor conditions on an open field. After temporally aligning the EKF and MCS data, the takeoff and landing phases were excluded, and a representative contiguous $10\,\mathrm{s}$ hover interval from each flight was extracted for analysis.
\subsection{Uncertainty Quantification}

The uncertainty in UAV positioning is characterized using a synchronized dataset obtained by comparing EKF-based state estimates with reference measurements provided by the MCS. Following the \textit{Guide to the Expression of Uncertainty in Measurement (GUM)}~\cite[Sec.~4.2]{JCGMGUM}, the deviations between EKF estimates and reference positions are used to statistically quantify the positioning error. The resulting uncertainty parameters in the $x$-, $y$-, and $z$-directions are computed from the empirical error statistics and summarized in Table~\ref{pos_uncert_tab}~\cite{Pos_uncert}. These values are directly used as input to the subsequent uncertainty propagation analysis.

\begin{table}[]
\centering
\begin{tabular}{lccc}
\hline
\textbf{GUM parameter} & \textbf{$x$-direction} & \textbf{$y$-direction} & \textbf{$z$-direction} \\
\hline
$\bar{q}^{e}_{\text{UAV}}$ & -0.2 m & 0.08 m & -0.61 m \\[2pt]
$u(\bar{q}^{e}_{\text{UAV}})$ & 0.004 m & 0.004 m & 0.004 m \\[2pt]
$u(q_{\text{UAV}})$ & 0.40 m & 0.40 m & 0.38 m \\
\hline
\end{tabular}
\caption{GUM-based uncertainty parameters for EKF-based UAV position estimation}
\label{pos_uncert_tab}
\end{table}

\section{Uncertainty Propagation Framework}
We establish a systematic link between UAV position uncertainty and the RIS-assisted channel by propagating EKF-based state uncertainties through the geometric and signal-level transformations. While the EKF provides estimates and uncertainties in the physical domain, these are first mapped to the Tx--RIS and RIS--Rx distances $d_m^h$ and $d_m^g$, where classical GUM-based moment propagation applies directly. In contrast, the subsequent channel construction involves coherent sums of complex-valued contributions, where phases depend nonlinearly on distance through $\exp\!\left(-j\frac{2\pi}{\lambda}(d_m^h + d_m^g)\right)$. As a result, linear uncertainty propagation is only valid under small phase perturbations, as higher-order effects of the exponential mapping become non-negligible otherwise.

Our previous work~\cite{prop_orient} operated in this small-variance regime using first-order Jacobian linearization. In this paper, we go beyond that assumption by directly propagating statistical moments through the full nonlinear complex-valued channel expression. This captures higher-order phase uncertainty effects and provides a consistent mapping from geometric uncertainty to the effective channel $h^{\mathsf{eff}}$.
\subsection{GUM-Compliant Moment-based Uncertainty Propagation}

Given an arbitrary measurement model \vspace{-0.1cm}
\begin{align}\label{eq:measFun}
g = f(q_1,q_2,\dots,q_N),
\end{align}
the quantity of interest $g$, referred to as the measurand, is modeled as a deterministic function of the input (influence) quantities $q_1, q_2, \dots, q_N$. According to the GUM framework~\cite[Sec.~5]{JCGMGUM}, the law of propagation of uncertainty yields the combined standard uncertainty
\vspace{-0.1cm}
\begin{align}\label{GumProp}
u^2(g)
=
\sum_{i=1}^{N} c_i^2\, u^2(q_i)
+
2 \sum_{i=1}^{N-1} \sum_{j=i+1}^{N}
c_i c_j\, u(q_i,q_j),
\end{align}
where $c_i = \frac{\partial f}{\partial q_i}$ denotes the sensitivity coefficient with respect to $q_i$, and $u(q_i,q_j)$ represents the covariance between $q_i$ and $q_j$. In the special case of mutually uncorrelated inputs, the covariance terms vanish.

\subsection{Uncertainty Propagation to Amplitude and Phase}

Based on the previously introduced moment-based propagation framework, we now propagate the UAV position uncertainty over the geometric quantities into the channel's amplitude and phase values. To this end, let $\mathbf{p}_\mathsf{UAV} = [p_{\mathsf{UAV},x},p_{\mathsf{UAV},y},p_{\mathsf{UAV},z}]^T$ denote the UAV center of gravity (COG), $\mathbf{r}_m$ the position vector of RIS element $m$ relative to the UAV's COG, and $\mathbf{R}(\Phi,\Theta,\Psi)$ the combined three-dimensional rotation matrix. The latter is described by the Euler angles roll $\Phi$, pitch $\Theta$, and yaw $\Psi$, corresponding to rotations around the $x$-, $y$-, and $z$-axes, respectively. Under the assumption that the UAV rotates only around its COG, the position of RIS element $m$ is given by \vspace{-0.1cm}
\begin{align}\label{eq:Pm}
  \mathbf{p}_m(x,y,z)
  =
  \mathbf{p}_\mathsf{UAV}
  +
  \mathbf{R}(\Phi,\Theta,\Psi)\mathbf{r}_m.\\[-15pt]\nonumber
\end{align}
Given the fixed transmitter position $\mathbf{p}_\mathsf{Tx}$, the Tx--RIS distance for element $m$ can be expressed as \vspace{-0.05cm}
\begin{align}\label{dist}
  d^h_m
  =
  \norm{\mathbf{p}_m(x,y,z)-\mathbf{p}_\mathsf{Tx}},\\[-15pt]\nonumber
\end{align}
which can be directly interpreted as a real-valued measurement function of the form introduced in (\ref{eq:measFun}).

This reformulation enables the application of the GUM-based moment framework developed in the previous section by calculating the sensitivity coefficients as \vspace{-0.1cm}
\begin{align}\label{sensD}
\hspace{-0.15cm} c_q^{d_m^h} \hspace{-0.075cm}=\hspace{-0.075cm}  \frac{\partial d_m^h}{\partial q}  \hspace{-0.05cm}\overset{(\ref{dist})}{=}\hspace{-0.05cm}
\frac{p_{m,q} - p_{\mathsf{Tx},q}}
{\|\mathbf{p}_m(x,y,z) - \mathbf{p}_\mathsf{Tx}\|}, \quad \forall q\in\{x,y,z\}.
\end{align}
The derivation of the RIS–Rx distance $d_m^g$ and its corresponding sensitivity coefficients follows an analogous procedure to that presented above for the Tx–RIS case and is therefore omitted for brevity.

%
With the above definitions, the standard combined uncertainties of the distances can be written as
\begin{align}\label{d_uncert}
  u^2(d_m^h) = \hspace{-0.2cm}\sum_{q\in\{x,y,z\}}\hspace{-0.1cm} (c_q^{d_m^h})^2 u^2(q),\,\,
  u^2(d_m^g) =\hspace{-0.1cm} \hspace{-0.2cm}\sum_{q\in\{x,y,z\}} \hspace{-0.1cm} (c_q^{d_m^g})^2 u^2(q) \\[-20pt]\nonumber
\end{align}
Although these expressions describe the marginal uncertainties of $d_m^h$ and $d_m^g$, both quantities depend on the same underlying UAV state, which induces statistical coupling that must be accounted for within the GUM framework to avoid biased uncertainty estimates~\cite[Sec.~5.2]{JCGMGUM}. The covariance is therefore obtained via the law of propagation of uncertainty for correlated inputs~\cite[Sec.~5.2]{JCGMGUM}, yielding \vspace{-0.0cm}
\begin{align}\label{u_dh_dg}
u(d_m^h,d_m^g)
= \sum_{q\in\{x,y,z\}}
c_q^{d_m^h}\, c_q^{d_m^g}\, u^2(q),\\[-20pt]\nonumber
\end{align}
where the result follows under the assumption that the position components $q \in \{x,y,z\}$ are mutually independent.

Building upon the distance uncertainty analysis, the uncertainty is subsequently propagated to the amplitude and phase components of the cascaded Tx--RIS--Rx channel associated with reflecting element $m$.
To this end, the cascaded channel coefficient in (\ref{h_casc}) is reformulated as
\begin{align}\label{h_casc_prop}
  h_m^\mathsf{casc} =
  \frac{c^2}{(4 \pi \nu)^2 d_m^h d_m^g}
  e^{j \frac{2\pi}{\lambda}(d_m^h+d_m^g)}
  = A_m e^{j\,P_m},
\end{align}
from which the amplitude and phase components are directly obtained. This representation allows us to apply the same GUM-based uncertainty propagation framework as detailed in \cite{prop_orient}. In particular, by explicitly accounting for the covariance between $d_m^h$ and $d_m^g$ and utilizing \eqref{GumProp} the respective combined uncertainties are given by
\begin{align}
u^2(A_m)
=&
A_m^2
\left(
\frac{u^2(d_m^h)}{(d_m^h)^2}
+
\frac{u^2(d_m^g)}{(d_m^g)^2}
+
2\frac{u(d_m^h,d_m^g)}{d_m^h d_m^g}
\right),
\end{align}
\begin{align}
u^2(P_m)
=&
\left(\frac{2\pi}{\lambda}\right)^2
\left(
u^2(d_m^h)
+
u^2(d_m^g)
+
2u(d_m^h,d_m^g)
\right).
\end{align}

\subsection{Stochastic Uncertainty Propagation of the Cascaded RIS Channel}
Motivated by the nonlinear mapping $e^{jP_m}$ in \eqref{h_casc_prop}, we extend the propagation framework to the complex domain. To enable a tractable analysis, we interpret the previously obtained uncertainties $u^2(A_m)$ and $u^2(P_m)$ as variances of an equivalent Gaussian model for amplitude and phase fluctuations. Accordingly, we model \vspace{-0.15cm}
\begin{align}
A_m &\sim \mathcal{N}(\mu_{A_m}, u^2(A_m)),\\
P_m &\sim \mathcal{N}(\mu_{P_m}, u^2(P_m)),
\end{align}
where $\mu_{A_m}$ and $\mu_{P_m}$ are obtained from the EKF estimates.

Given this stochastic representation, we first evaluate the mean effect of the nonlinear phase term. In particular, using the characteristic function of a Gaussian random variable, the expectation of the complex exponential becomes \vspace{-0.05cm}
\begin{align}
\mathbb{E}[e^{jP_m}] = e^{j\mu_{P_m}} e^{-u^2(P_m)/2},
\end{align}
where $e^{-u^2(P_m)/2}$ acts as a coherence attenuation factor. This directly yields the mean cascaded channel \vspace{-0.05cm}
\begin{align}
\mathbb{E}[h_m^\mathsf{casc}] = \mu_{A_m} e^{j\mu_{P_m}} e^{-u^2(P_m)/2}.
\end{align}

\begin{figure}
\centering
\includegraphics[width=0.65\linewidth]{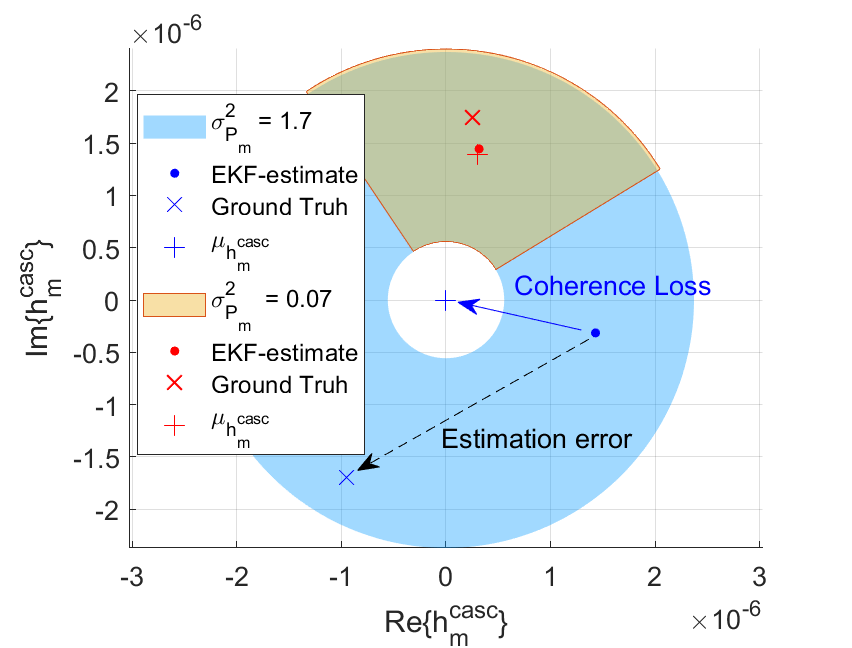}
\caption{Illustration of the EKF-estimated cascaded channel $h_m^\mathsf{casc}$, its mean value $\mu_{h_m^\mathsf{casc}}$, and the corresponding coherence loss induced by phase uncertainty for different values of $u^2(P_m)$, including the associated 95\% confidence intervals.}
\label{fig:cohLoss_perM}
\end{figure}

We next characterize the second-order statistics of $h_m^\mathsf{casc}$. A direct application of standard GUM propagation is not feasible due to the nonlinear complex-valued mapping. We therefore instead use an equivalent moment-based representation
\begin{align}
u^2(h_m^\mathsf{casc})
= \mathbb{E}[|h_m^\mathsf{casc}|^2]
- |\mathbb{E}[h_m^\mathsf{casc}]|^2,
\end{align}
using the second-moment decomposition $\mathbb{E}[X^2] = \mathbb{E}[X]^2 + \mathrm{Var}(X)$, which separates the total power into its mean and fluctuation components. Here, the first term simplifies using $|e^{jP_m}|=1$, yielding \vspace{-0.05cm}
\begin{align}
\mathbb{E}[|h_m^\mathsf{casc}|^2] = \mu_{A_m}^2 + u^2(A_m),
\end{align}
while the second term follows directly from the previously derived mean. Combining both contributions results in
\begin{align}
u^2(h_m^\mathsf{casc})=u^2(A_m)+\mu_{A_m}^2\left(1-e^{-u^2(P_m)}\right),
\end{align}
which shows that phase uncertainty introduces an additional variance term due to coherence loss.
To illustrate the difference between the EKF-estimated cascaded channel $h_m^\mathsf{casc}$ and its mean $\mu_{h_m^\mathsf{casc}}$, as well as the impact of phase uncertainty, we consider EKF estimates, ground-truth values, and 95\% confidence intervals for amplitude and phase under two representative values of $u^2(P_m)$ in Fig.~\ref{fig:cohLoss_perM}.

For small phase variance, $P_m$ remains tightly concentrated around its mean, resulting in nearly aligned phases and predominantly coherent combining. Consequently, $\mu_{h_m^\mathsf{casc}}$ closely follows the deterministic case, as shown by the orange markers and confidence region in Fig.~\ref{fig:cohLoss_perM}. In contrast, for large phase variance, the wrapped phase $e^{jP_m}$ becomes approximately uniformly distributed on the unit circle due to $2\pi$ periodicity. This induces phase decorrelation across realizations and drives $\mu_{h_m^\mathsf{casc}}$ toward zero, corresponding to a complete loss of coherence (blue markers).

\subsection{Uncertainty Propagation to Effective RIS Channel}
To further propagate the previously obtained amplitude and phase uncertainties through the RIS phase control, we express the effective cascaded channel for the $m$-th element as
\begin{align}
h_m^\mathsf{eff} = A_m e^{j(P_m + \varphi_m)},
\end{align}
where $\varphi_m$ is deterministic. Similarly to the procedure for $h_m^\mathsf{casc}$, the phase $P_m$ and amplitude $A_m$ can be treated as stochastic quantities with means $\mu_{P_m}$ and $\mu_{A_m}$ and variances $u^2(P_m)$ and $u^2(A_m)$, respectively. The mean of the effective channel then follows as
\begin{align}
\mu_{h_m^\mathsf{eff}}
&= \mu_{A_m} \, e^{j(\mu_{P_m} + \varphi_m)} e^{-u^2(P_m)/2},
\end{align}
and since $|e^{j(\cdot)}| = 1$, the second-order moment becomes
\begin{align}\label{sec_ord_mom}
\mathbb{E}[|h_m^\mathsf{eff}|^2] = \mu_{A_m}^2 + u^2(A_m),
\end{align}
Combining the first- and second-order moments yields
\begin{align}\label{u2_h_m^eff}
u^2(h_m^\mathsf{eff})
&= u^2(A_m)
 + \mu_{A_m}^2 \left(1 - e^{-u^2(P_m)}\right).
\end{align}
In order to extend these results to the overall RIS channel, we can write the effective cascaded response as
\begin{align}
  h^\mathsf{eff} = \sum_{m=1}^M h_m^\mathsf{eff}.
\end{align}
Here, assuming statistical independence between reflecting elements, the GUM covariance propagation reduces to a summation of individual contributions. The mean effective channel is therefore\vspace{-0.1cm}
\begin{align}\label{mu_h^eff}
\mu_{h^\mathsf{eff}}
=
\sum_{m=1}^M
\mu_{A_m} e^{j(\mu_{P_m}+\varphi_m)} e^{-u^2(P_m)/2},
\end{align}
while the total variance follows as\vspace{-0.1cm}
\begin{align} \label{u2_h^eff}
u^2(h^\mathsf{eff})
=
\sum_{m=1}^M u^2(h_m^\mathsf{eff}),
\end{align}
which is consistent with the law of propagation of uncertainty for uncorrelated complex-valued contributions.

\section{System Design Implications Under Uncertainty}
%
The results in \eqref{mu_h^eff} and \eqref{u2_h^eff} provide an interpretable decomposition of uncertainty propagation through each RIS element. The variance of the per-element effective channel $h_m^\mathsf{eff}$ in \eqref{u2_h_m^eff} consists of intrinsic amplitude uncertainty $u^2(A_m)$ and an additional phase-induced term, which scales with the squared mean amplitude and includes an exponential attenuation factor $e^{-u^2(P_m)}$. This highlights that phase uncertainty not only introduces random phase rotations but also reduces the effective coherent gain.

Importantly, the deterministic RIS phase shift $\varphi_m$ does not affect the statistical dispersion, as $u^2(h_m^\mathsf{casc}) = u^2(h_m^\mathsf{eff})$. Instead, it only governs the phase alignment of contributions across elements, thereby affecting the mean response $\mu_{h_m^\mathsf{eff}}$ but leaving second-order statistics unchanged. For the overall RIS-assisted channel, these effects aggregate across elements, yielding a tractable large-scale characterization that enables design insights and performance bounds, as discussed in this section.

\subsection{Scaling Law of RIS Coherence Under Uncertainty}

Utilizing the propagated statistics of $h^\mathsf{eff}$, we are able to characterize how uncertainty scales with the number of reflecting elements. In the special case of identical elements, i.e., $\mu_{A_m}=\mu_A$ and $u^2(P_m)=u^2(P)$, the scaling laws simplify to
\begin{align}
\mu_{h^\mathsf{eff}} &\sim M \mu_A e^{-u^2(P)/2}, \\
u^2(h^\mathsf{eff}) &\sim M \left[u^2(A) + \mu_A^2 \left(1 - e^{-u^2(P)}\right)\right],
\end{align}
showing that both the mean and variance grow linearly with the number of reflecting elements $M$. However, the phase uncertainty introduces a fundamental coherence loss through the exponential attenuation factor, which prevents the ideal quadratic power scaling associated with perfectly phase-aligned RIS arrays. Instead, the achievable gain is reduced by an uncertainty-dependent coherence factor, which becomes the dominant performance limitation in large-scale RIS setups.

To quantify the impact of phase uncertainty on RIS performance, we consider the expected received power $\mathbb{E}[|h^\mathsf{eff}|^2]$ as a basis for deriving performance bounds. Under the assumption of independent phase perturbations and optimal phase alignment $\varphi_m = -\mu_{P_m}$, the mean effective channel simplifies to \vspace{-0.1cm}
\begin{align}
  \mathbb{E}[h^\mathsf{eff}] =
  \sum_{m=1}^M \mu_{A_m} e^{-\sigma_{P_m}^2/2}. \\[-20pt]\nonumber
\end{align}
Similar to \eqref{sec_ord_mom}, we can utilize the second-moment identity to decompose the expected power into coherent and incoherent contributions as\vspace{-0.1cm}
\begin{align}\label{eq:e_pow_analytical}
  \mathbb{E}[|h^\mathsf{eff}|^2]
  =
  \Big(\sum_{m=1}^M \mu_{A_m} e^{-\sigma_{P_m}^2/2}\Big)^2
  +
  \sum_{m=1}^M \mu_{A_m}^2 (1 - e^{-\sigma_{P_m}^2}).\\[-29pt] \nonumber
\end{align}
resulting in the bounds\vspace{-0.1cm}
\begin{align}
  \sum_{m=1}^M \mu_{A_m}^2
  \;\le\;
  \mathbb{E}[|h^\mathsf{eff}|^2]
  \;\le\;
  \left(\sum_{m=1}^M \mu_{A_m}\right)^2,
\end{align}
which correspond to fully incoherent and fully coherent combining, respectively. For the special case of equal amplitudes $\mu_{A_m} = \mu_{A}$ and identical phase variance $u^2(P_m) = u^2(P)$, the expected power simplifies to
\begin{align}\label{anal_bound}
  \mathbb{E}[|h^\mathsf{eff}|^2] = \mu_{A}^2 \left(
  M^2 e^{-u^2(P)}
  +
  M(1 - e^{-u^2(P)})
  \right).
\end{align}
For RIS beamforming to provide a meaningful gain, the coherent contribution must dominate, i.e., $M^2 e^{-u^2(P)} \gtrsim M$,
resulting in the condition
\begin{align}
u^2(P) \lesssim \ln M,
\end{align}
which defines the maximum tolerable phase uncertainty for preserving coherent array gain. In other words, to prevent coherence loss from dominating as the array grows, the allowed phase uncertainty must scale only logarithmically with the number of reflecting elements. However, while this condition ensures feasibility of coherent scaling, it does not fully capture the gradual degradation of performance observed in practical regimes.
\subsection{Performance-Driven Coherence Threshold}
A more intuitive characterization of coherence can therefore be obtained by defining a relative performance threshold. Instead of distinguishing between coherent and incoherent scaling, we require that the RIS preserves a fixed fraction $\eta_0 \in (0,1]$ of the ideal coherent gain $M^2$. This leads to the definition \vspace{-0.1cm}
\begin{align}
\frac{\mathbb{E}[|h^\mathsf{eff}|^2]}{M^2} \ge \eta_0,
\end{align}
which provides a direct measure of usable coherent performance. By substituting the exact expression for the average power we obtain \vspace{-0.1cm}
\begin{align}
e^{-u^2(P)} + \frac{1}{M}\left(1 - e^{-u^2(P)}\right) \ge \eta_0
\end{align}
and rearranging for the phase variance gives the exact finite-$M$ performance-driven coherence threshold (PCT) \vspace{-0.1cm}
\begin{align}
u^2(P) \le -\ln\!\left(\frac{\eta_0 - \frac{1}{M}}{1 - \frac{1}{M}}\right),
\end{align}
which explicitly captures the interplay between phase uncertainty and array size. For the special case of a $-3$ dB threshold, i.e., losing half of the channel gain due to incoherent alignment, which corresponds to $\eta_0 \approx 0.5$, this reduces to \vspace{0.1cm}
\begin{align}\label{coh_bound}
u^2(P) \le -\ln\!\Big(\frac{0.5 - \frac{1}{M}}{1 - \frac{1}{M}}\Big).
\end{align}
This expression shows that the coherence boundary is slightly dependent on the array size for finite $M$, and only converges to $u^2(P) = \ln 2$ in the large-array limit.

\section{Numerical and Experimental Analysis}
For the analytical and measurement-based evaluation, the RIS prototype dimensions and UAV mounting configuration from the experimental setup are adopted. In particular, the carrier frequency is set to $\nu=5.376\,\mathrm{GHz}$, the RIS dimensions are chosen as $0.20\,\mathrm{m}\times0.16\,\mathrm{m}$, and the UAV mounting offset is given by $-0.30\,\mathrm{m}$ along the $z-$axis. The transmitter and receiver positions are defined as $\mathbf{p}_{\mathsf{Tx}}=[0,0,0.1]^T$ and $\mathbf{p}_{\mathsf{Rx}}=[5,0,0.1]^T$, respectively.

Further, the UAV position is selected as the nominal operating point of the measurement dataset, i.e., $(x,y,z)=(0,0,3.5)$, corresponding to a position directly above the transmitter \cite{Pos_uncert}. To isolate the influence of the RIS size on the propagated uncertainty and coherence loss, the physical RIS dimensions are kept fixed throughout the analysis, while the number of reflecting elements is varied accordingly. In addition, the UAV orientation is assumed to be constant with roll $\Phi$, pitch $\Theta$, and yaw $\Psi$ angles set to $(0,0,0)$ during the entire evaluation with the transmitter and receiver antenna gains normalized to $G_T=G_R=1$.
\subsection{Tightness Analysis of the Performance-Driven Coherence Threshold}
\begin{figure}
\begin{minipage}{0.49\linewidth}
    \centering
    \includegraphics[width=\linewidth]{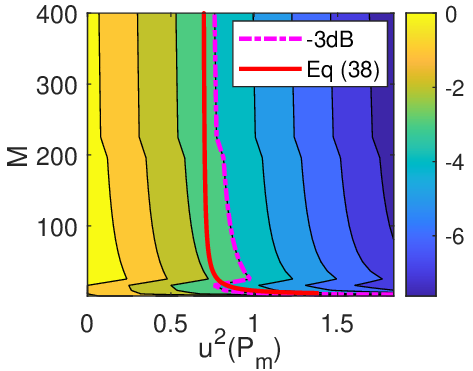}
\end{minipage}
\hfill
\begin{minipage}{0.49\linewidth}
    \centering
    \includegraphics[width=\linewidth]{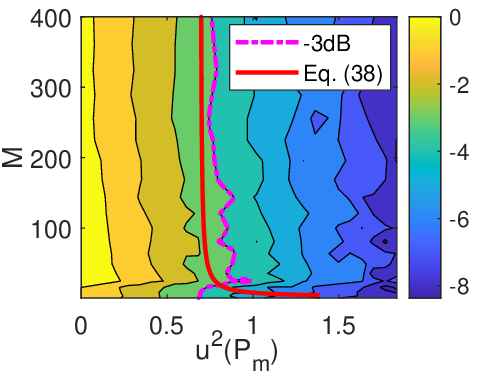}
\end{minipage}
\caption{$-3$ dB coherence bound \eqref{coh_bound} (red solid) and actual threshold (purple dashed): scaling-law-based analytical results \eqref{anal_bound} (left) and UAV-distribution-based Monte Carlo results (right), shown over phase uncertainty $u^2(P_m)$ and number of RIS elements $M$.}
\label{fig:PDCT}
\vspace{-0.1cm}
\end{figure}

\begin{figure*}[t]
\centering

\begin{minipage}{0.31\linewidth}
    \centering
    \includegraphics[width=0.89\linewidth]{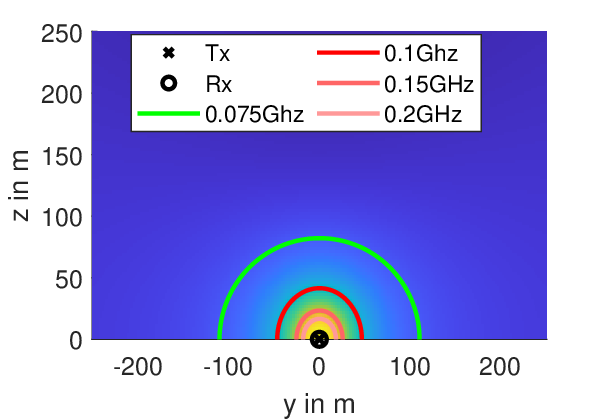}
\end{minipage}
\hfill
\begin{minipage}{0.31\linewidth}
    \centering
    \includegraphics[width=0.89\linewidth]{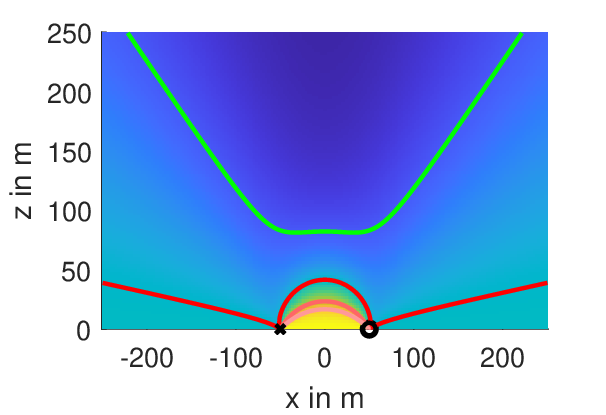}
\end{minipage}
\hfill
\begin{minipage}{0.33\linewidth}
    \centering
    \includegraphics[width=0.89\linewidth]{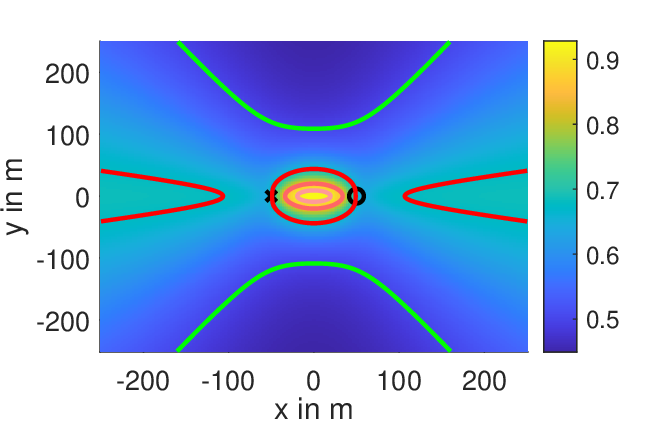}
\end{minipage}
\caption{Visualization of UAV-mounted RIS's feasible deployment area in the $yz$-, $xz$- and $xy$-planes, with $z = 15\,\mathrm{m}$ for the $xy$-plane.}
\label{fig:uav_area}
\end{figure*}
We analyze the tightness of the performance-driven coherence threshold (PCT) in \eqref{coh_bound} as a function of the phase uncertainty $u^2(P_m)$ and the number of reflecting elements $M$. To this end, we consider two complementary evaluation cases in Fig.~\ref{fig:PDCT}: an analytical scaling law (Case I) based on propagated second-order statistics, and a Monte Carlo evaluation (Case II) based on experimentally derived uncertainty distributions.

In Case I, the UAV position is fixed at the nominal EKF operating point, while different uncertainty levels in $(x,y,z)$ are propagated through the model to obtain $u^2(P_m)$. A scalar phase uncertainty is then obtained by aggregating the component-wise contributions, yielding a deterministic mapping between $(M, u^2(P_m))$ and the coherence bound. As shown in Fig.~\ref{fig:PDCT}, the PCT accurately captures the analytical threshold behavior, including the asymptotic scaling for small and large $M$, with deviations mainly in the intermediate regime where the coherence transition becomes pronounced.

In Case II, 1000 UAV position realizations are generated around the nominal EKF estimate using a covariance model derived from the measured uncertainty ratios in $(x,y,z)$. Each realization is propagated through the channel model to obtain samples of $A_m$, $P_m$, and $h^{\mathsf{eff}}$, enabling a measurement-consistent evaluation of coherence behavior under stochastic uncertainty.
Compared to Case I, the Monte Carlo results exhibit increased variability due to realization-dependent channel conditions and element-wise uncertainty propagation. Nevertheless, the PCT remains accurate in capturing the asymptotic $-3\,\mathrm{dB}$ threshold behavior as $M \rightarrow \infty$. For small arrays, fluctuations in individual elements lead to higher variability in the effective channel, whereas larger arrays exhibit self-averaging, resulting in more stable coherence behavior. Consequently, the threshold is most accurate in the large-$M$ regime and degrades mainly for small $M$ and high phase uncertainty.
\subsection{Feasible Deployment Area}
Based on the validated tightness of the proposed PCT, we derive feasible UAV-mounted RIS deployment regions that guarantee a desired coherence threshold. The resulting operating region depends on the target threshold, RIS size, Tx/Rx geometry, positioning uncertainty, and carrier frequency.

Fig.~\ref{fig:uav_area} illustrates the feasible regions in the $yz$-, $xz$-, and $xy$-planes (for $z=15\,\mathrm{m}$), using the experimental RIS setup with $M=120$ elements. The transmitter and receiver are located at $\mathbf{p}_{\mathsf{Tx}}=[-50,0,0.1]^\top$ and $\mathbf{p}_{\mathsf{Rx}}=[50,0,0.1]^\top$, respectively, and the uncertainty parameters are taken from \cite{Pos_uncert} (Table~\ref{pos_uncert_tab}). The feasible regions are evaluated for carrier frequencies between $0.075\,\mathrm{GHz}$ and $0.2\,\mathrm{GHz}$ under the $-3$~dB condition in \eqref{coh_bound}.

The results show that the largest coherence is achieved when the UAV-mounted RIS is positioned near the Tx–Rx midpoint, where the effective propagation distance is minimized. Lower carrier frequencies significantly enlarge the feasible region due to reduced phase sensitivity to uncertainty. However, the results also reveal that midpoint placement is not always optimal for all heights, as more robust regions emerge at moderate lateral offsets along the Tx–Rx axis. This effect becomes more pronounced at lower frequencies, where the feasible region transitions into a broader, non-uniform shape, indicating that off-center UAV-mounted RIS placement can outperform midpoint positioning in certain uncertainty regimes.

\section{Conclusion}

This paper has analyzed the impact of UAV positioning uncertainty on the coherence behavior of RIS-assisted channels using a GUM-consistent propagation framework. A closed-form stochastic model for the cascaded channel was derived, capturing coherence loss due to nonlinear phase uncertainty effects. Based on this model, a performance-driven coherence threshold was introduced and validated via an analytical scaling law and a measurement-informed Monte Carlo evaluation. The results show strong agreement in the asymptotic regime, while deviations occur for small RIS sizes and high uncertainty due to limited self-averaging. Using the PCT for deployment analysis reveals that the Tx-Rx midpoint is not always optimal, with off-center UAV positions becoming favorable when operating at increased heights. \vspace{-0.2cm}

\balance
\bibliographystyle{IEEEtran}
\small
\vspace{-0.0cm}
\bibliography{bibliography}
\balance
\end{document}